\documentclass[twocolumn,superscriptaddress,showpacs,amsmath,amssymb,prl,longbibliography,floatfix]{revtex4-1}

\usepackage{graphicx} 
\usepackage{dcolumn}  
\usepackage{bm}       
\usepackage[usenames,dvipsnames]{color}
\definecolor{URLCOL}{rgb}{0,0.52,0.83} 
\definecolor{LINKCOL}{rgb}{0.05,0.5,0} 
\definecolor{orange}{rgb}{0.6,0.3,0} 
\definecolor{CITECOL}{rgb}{0.25,0,0.48} 
\usepackage{epstopdf}

\usepackage[pdftex,bookmarks,breaklinks,bookmarksopen,bookmarksnumbered,colorlinks,linkcolor=LINKCOL,linktocpage,citecolor=CITECOL,urlcolor=URLCOL,pdfpagemode=UseOutline,pdftex]{}

\usepackage{fancyhdr}
\fancypagestyle{titlepage}
{\chead{file: \jobname}\rhead{Total pages: \pageref{page:end}}}

\usepackage{soul}
\usepackage{booktabs}
\usepackage{natbib}

\definecolor{TITLECOL}{rgb}{0.1,0.2,0.7} 
\definecolor{SECOL}{rgb}{0.1,0.2,0.7} 
\definecolor{CONTENTSCOL}{rgb}{0.1,0.2,0.7} 
\definecolor{SSECOL}{rgb}{0.25,0,0.48} 
\definecolor{SSSECOL}{rgb}{0.2,0.08,0.53} 
\definecolor{FINCOL}{rgb}{0.01,0.3,0.07} 

\def\coloredtitle#1{\title{\textcolor{TITLECOL}{#1}}} 
\def\coloredauthor#1{\author{\textcolor{CITECOL}{#1}}} 

\definecolor{URLCOL}{rgb}{0,0.17,0.43} 
\definecolor{LINKCOL}{rgb}{0.05,0.4,0} 
\definecolor{CITECOL}{rgb}{0.35,0,0.48} 

\def\sss{\scriptscriptstyle\rm}

\def\bea{\begin{eqnarray}}
\def\eea{\end{eqnarray}}
\def\ben{\begin{equation}}
\def\een{\end{equation}}
\def\benu{\begin{enumerate}}
\def\enu{\end{enumerate}}

\def\bei{\begin{itemize}}
\def\eei{\end{itemize}}
\def\beit{\begin{itemize}}
\def\eit{\end{itemize}}
\def\benu{\begin{enumerate}}
\def\enu{\end{enumerate}}

\def\br{{\bf r}}

\def\s{_{\sss S}}

\def\Hxc{_{\sss HXC}}

\def\n{n}

\begin{document}

\coloredtitle{
Can ensemble density functional theory yield accurate double excitations?
}
\coloredauthor{Francisca Sagredo}
\affiliation{Department of Chemistry, University of California, Irvine, CA 92697}
\coloredauthor{Kieron Burke}
\affiliation{Department of Chemistry, University of California, Irvine, CA 92697}
\affiliation{Department of Physics and Astronomy, University of California, Irvine, CA 92697}
\date{\today}

\begin{abstract}
The recent use of a new ensemble in density functional theory (DFT) to
yield direct corrections to the Kohn-Sham transitions yields
the elusive double excitations that are missed by time-dependent DFT with
the standard adiabatic approximation.  But accuracies
are lower than for single excitations, and formal arguments
suggest that direct corrections at the exchange level should not be sufficient.  
We show that in principle, EDFT
with direct corrections can yield accurate doubles and explain the error
in naive formal arguments about TDDFT.
Exact calculations and analytic results
on a simple model, the Hubbard dimer, illustrate the results, showing that
the answer to the title question is typically yes.

\end{abstract}

\maketitle


Time-dependent density functional theory (TDDFT) is a popular tool for
calculating electronic excitations\cite{RG84,C96,MMNG12,U12}, but with current approximations, has
some severe limitations. Within the adiabatic
approximation used in almost all practical calculations and all standard
codes, double (and multiple) excitations are entirely missed by TDDFT\cite{M16}.  While in some
cases these can be recovered in an ad-hoc fashion using dressed TDDFT,
which approximates the frequency dependence, there is no general procedure
for capturing these relevant excitations.

On the other hand, ensemble DFT (EDFT) is almost as venerable, but is much less
used\cite{GOK88,GOKb88,GOKc88}.  Unlike TDDFT, which employs linear response around the ground-state
to deduce excitation energies, EDFT is based on a variational theorem for a decreasing set of eigenstates,
from which transition frequencies can be deduced.  Using the original ensemble
of Gross, Oliveria, and Kohn (GOK) \cite{GOK88}, there has been much formal progress over
three decades, but accurate approximations have been difficult to develop.
An important step forward came with the identification of ghost-interaction errors,
and their removal in the work of Papaconstantinou, Gidopoulos,  and Gross\cite{GPG02}, and in using  the symmetry-adapted Hartree-exchange\cite{YTPB14, PYTB14}, now referred to as the ensemble exact exchange (EEXX)\cite{GP17}. Furthermore, new work in the generalized adiabatic connection, and the investigation of charge transfer within EDFT \cite{FF14,GKP17}, as well other recent contributions\cite{Nb95,Nb98,TN03,AKF16} have all been important to push EDFT forward. But these recent publications rarely focus on double excitations, except in Ref.\cite{YPBU17}.

In Ref.\cite{YPBU17}, an alternative ensemble (GOK II) was examined, which has several formal and
practical advantages. Moreover, using the exact-exchange approximation above,
and taking the weights of excited states to (almost) zero, Ref.\cite{YPBU17} found
a simple direct ensemble correction (DEC) to Kohn-Sham transition frequencies, analogous to 
expressions in TDDFT.  Preliminary tests on atoms and a simple model (1D Hooke's atom)
showed that, for single excitations, results were comparable to or better than
standard TDDFT results.  More importantly, double excitations were predicted by the
new method, but substantially
less accurately than for singles in every case.

The present work addresses the question:  Does the DEC method of Ref.\cite{YPBU17}
really produce a useful path toward calculating double excitations, or are their
results more-or-less accidental, in the sense that there is no limit in which the
method yields the right corrections to KS transitions?  There are formal reasons
for raising these concerns. In TDDFT correlation contributions to the XC kernel are needed to produce the frequency
dependence required to generate double excitations.  But the exchange correction used in DEC
does not include correlation.  So how can it yield accurate corrections?

\begin{figure}[!htp]
  \includegraphics[width=\linewidth]{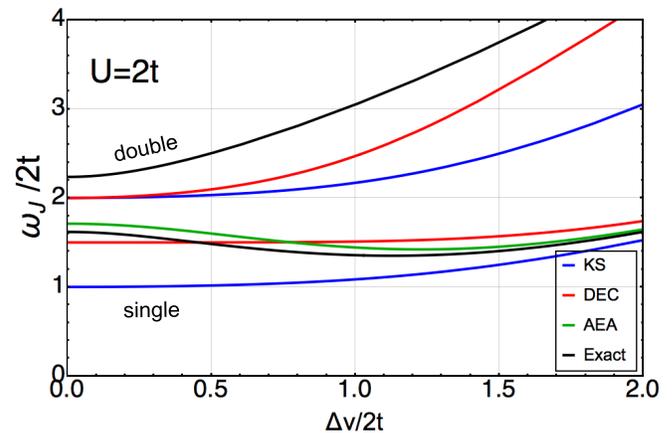}
  \caption{Transition frequencies versus onsite potential difference for the weakly correlated Hubbard dimer at with $2t=1$. The exact many body solution (black) for single (bottom four curves) and double excitation (top three curves) are compared against the Kohn Sham (KS), Adiabatically Exact Approximation (AEA), and Direct Ensemble Correction (DEC). }
  \label{boat:2vAE_short}
\end{figure}
We answer these questions with calculations on a simple model, the asymmetric Hubbard
dimer, which provides explicit analytic results.
Our principal results are shown in Fig. \ref{boat:2vAE_short}, and described in detail within.
While DEC in EDFT and adiabatic TDDFT both yield accurate results (but not everywhere) for the first single excitation,
only DEC makes a prediction for the double, and is typically accurate for weakly correlated systems.
We find a substantial exchange correction to the Kohn-Sham transition of the double-excitation.
We also explain the connection with TDDFT, and the relation among various expansions in 
powers of the coupling constant.  Finally, we explain why correlation is needed to find double (and multiple) excitations in TDDFT, but not in EDFT.

{\bf G{\"o}rling-Levy (GL) perturbation theory} \cite{GL95} is the appropriate tool for studying excitations in DFT
for weakly-correlated systems.  Expanding the energy of the $J-th$ many-body state in powers of $\lambda$, the electron-electron repulsion,
while keeping the density fixed (the adiabatic connection\cite{LP85,GL95}), one finds
\ben
E_J=\epsilon_J + \lambda \Delta v_{JJ} + \lambda^2\left(
\sum_{J'\neq J} \frac{|v_{J'J}|^2}{\epsilon_{J'}-\epsilon_J} - v_{c,J,J} \right),
\label{eq.1}
\een
where the KS energy of the $J-th$ state is $\epsilon_J$, $v_{JJ}$ is the expectation value of
the electron-electron repulsion operator minus the Hartree and exchange potentials \cite{ZB04},
and $v_{c,J,J}$ is the expectation of the 2nd-order correlation potential. Here we label excitations by the level of the excitation of the adiabatically connected KS determinant relative to the occupation of the KS ground state \cite{CFMB18}.

{\bf{TDDFT}} yields transition frequencies via linear response.  The exact density-density response function is
\ben
\chi(\br,\br',\omega)=\sum_{J\neq 0} \frac{m_J(\br)\, m^*_J(\br')}{\omega - \omega_{J} + i 0_+} +(c.c.),
\een
where $m_J(\br)=
\langle 0 | \hat \n(\br) | J\rangle$, $\hat \n(\br)$ is the density
operator, and transitions occur at its poles \cite{MMNG12}.  The KS counterpart is its value when $\lambda\to 0$, keeping the density fixed.  Then the
wavefunctions become single Slater determinants (typically), and the difference in the inverse of the
two response functions is called the Hartree-exchange-correlation kernel\cite{CFMB18}.  Because the density is a single-particle
operator, $m^{(0)}_J(\br)=0$
unless $J$ is a single-particle excitation, i.e., double excitations do not appear
in the KS response.  If the kernel is then approximated as frequency-independent (called the adiabatic approximation),
it does not affect the pole-structure, so the approximated response has only single excitations.  In the specific
case of two electrons whose ground-state is a singlet, the exchange kernel is static.  Thus, correlation effects
(at least second-order in $\lambda$) are needed in TDDFT to extract double excitations.  The approximate kernel of dressed
TDDFT, which applies only to doubles that are strongly coupled to singles, uses the square of the exchange
matrix element of the Hamiltonian, i.e., a second-order contribution \cite{MZCB04,CZMB04}. Thus, without
correlation contributions, TDDFT cannot typically produce double excitations.
\\

{\bf{Ensemble DFT}} is based on a variational principle for ensembles that are
a mixture of the lowest $M$ electronic eigenstates, for a chosen
set of weights ${\bf{w}} = \{{w_{J}}\}_{J=0...M-1}$ 
that are normalized and monotonically
decreasing.  Just as in ground-state DFT, one can define $F_{\bf w}[\n]$ which, when
added to the external potential and minimized over (ensemble) densities, yields the ensemble
energy \cite{GOK88,GOKb88,GOKc88}.  The GOK ensemble
has weight $w$ for the highest state, and all others chosen equal.
One can also define an ensemble KS system of non-interacting electrons by using the same ensemble and the correct minimizing ensemble
density. The change in $F_{\bf w}$ between interacting and non-interacting
defines the ensemble Hartree-exchange-correlation energy $E_{{\Hxc},\bf {w}}[\n]$, whose functional derivative
yields the corresponding contribution to the KS potential.

One complication of EDFT is that a range of values of $w$ is allowed (as long as normalization is possible), and the total energy of the system $E_{w}$ is exactly linear with respect to $w$, with its slope related to transitions of the system.  Almost all approximations lead to non-linear behavior with $w$, leading to different predictions depending on the value of $w$ chosen.
The (traditional)
Hartree energy, being quadratic in the density, has unphysical cross-terms proportional to $w_{J} w_{J'}$, 
which are referred to as ghost interaction errors.  
The careful removal of these errors from Hartree and exchange together yields greatly 
increased accuracy \cite{GPG02}.  Most recently,
this ensemble exact exchange (EEXX) \cite{GP17} has been shown to be the correct (energy-minimizing) choice to first-order in the interaction.

Ref.\cite{YPBU17} used an alternative ensemble suggested by GOK (called
GOK II), in which each state in the ensemble has weight $w$, except the ground state.  They also considered the limit as $w\to 0$, thereby using only the slope around $w=0$, yielding a unique answer that
is simply a correction to the ground-state KS transitions, i.e., there is no need to do an additional
self-consistent cycle for $w\neq 0$.  Finally, they also noted that, for the GOK II
ensemble, within EEXX, this direct energy correction requires only energy differences between the level
of interest and the ground state (and not all intervening states, as is otherwise
the case).  Plugging in the EEXX into the DEC approximation, and in the absence of degeneracies:
\ben
\Delta \omega^{EEXX}_{J} = \lambda (H_J-H_0)
\een
where $H_J$ is an exchange contribution depending only on the KS orbitals and energies of the $J$-th state(Eq.(9) of Ref.\cite{YPBU17}). They also calculated both single and double excitations
for a series of atoms and ions, and for the 1D Hooke's atom.  In all cases, the DEC/EEXX
yielded single excitations with accuracy comparable to that of TDDFT with standard approximations, 
while double excitations were also predicted, but with less accuracy.  The errors were ascribed to correlation effects missed by DEC/EEXX.

But the DEC/EEXX approximation of EDFT contains only a first order correction to the KS
transition frequencies, and so cannot generate corrections that are second-order
in the interaction (with fixed density).  So, are the first-order corrections
that it produces for multiple excitations trivial or meaningful?  And if meaningful,
how do such corrections arise in TDDFT?
\\
{\bf{Hubbard dimer}}:  The Hubbard model is a paradigm of strongly correlated physics, and
typically consists of an infinite lattice, with hopping and site-interaction terms \cite{H63}.  The
 dimer is likely the smallest meaningful model
of interacting fermions, with a Hilbert space of just 6 states\cite{CFSB15}.  It mimics strong correlation
effects of bond stretching, but is not a quantitatively accurate model for any first-principle Hamiltonian.  In its usual form, 
it is a simplified version of a minimal-basis model of two electrons on two atoms, with
one basis function per atom.  The Hamiltonian is
\ben
\hat{H} = -t\, \sum_{\sigma} \: (\hat{c}_{1\sigma}^{\dagger}\hat{c}_{2\sigma} + h.c) +
U \sum_{i} \hat{n}_{i,\uparrow\,}\hat{n}_{i,\downarrow} + \sum_i v_i \hat{n}_i  .
\label{eq:HH}
\een
Here $t$ is the electron hopping energy, $U$ is 
the the repulsion between the particles in each site, and the symmetry of the dimer is controlled by the 
potential difference, $\Delta v= v_{2}-v_{1}$, and the density is characterized by a single number, $\Delta n= n_{1}-n_{2}$. The Hubbard dimer is extremely useful 
for understanding ground-state DFT \cite{CFSB15}, 
especially when correlations are strong, and extensions and variations have been used in 
many time-dependent problems to understand TDDFT.  Its value comes 
from the ability to solve most problems analytically. A full discussion
of how linear-response TDDFT works for the dimer has just been completed\cite{CFMB18}.
Recently, the dimer has been used to illustrate EDFT weight-dependence\cite{DMSF18}, novel approaches to band
gaps\cite{SF18}, and 
approaches to noncollinear magnetism \cite{U18}.

Here, we study only singlet states, avoiding the complexities of spin-flipping transitions. There are then only 2 transitions, one to a single and one to a double excitation (the nature of a transition is determined by 
adiabatically turning off the interaction and labelling it based on its KS determinant). There are two parameters: $\Delta v$ measures the degree of asymmetry, while $U$ measures the strength of the interaction. When $\Delta v = U=0$, the dimer is a symmetric, tight-binding problem.  When $\Delta v/(2t)$ grows large,
the dimer is highly asymmetric, with both particles mostly on one site (in the ground state);  when $U/(2t)$ grows large, the
dimer has strong correlation effects, just like when a bond is stretched, and many ground-state
density functional approximations fail.  For $\Delta v=0$, the expansion about weak correlation diverges at
$U=4t$. Any the $\lambda$-dependence is found by replacing $U$ by $\lambda$ for a fixed $\Delta n$

Many of the most important results of this study appear in Fig. \ref{boat:2vAE_short}.  The solid black
lines are the well-known analytic results for the single (lower) and double (higher) excitations \cite{CFMB18}.
The value $U=2t$ is chosen to be a significant correlation strength, but still in the weakly-correlated
regime.  The blue  lines are the corresponding KS transitions, with the double at exactly twice
the single.  These are the exact KS transitions, meaning the transitions between occupied and unoccupied
KS orbitals of the exact ground-state KS potential, found from the functional derivative of the exact
ground-state XC energy\cite{CFSB15}.  

There are many lessons in this figure. As is typical for weakly
correlated systems, the KS transition frequencies are a reasonable zero-order approximation to
the exact optical excitations\cite{WGB05}.  We define $\Delta \omega_{J}=\omega_{J}-J\omega_{s}$ as the difference between the exact
and KS transition frequency.  We also note that the accuracy of the KS transitions is not uniform with $\Delta v$.  At $\Delta v=0$ (the symmetric case), $\Delta \omega_2 < \Delta \omega_1$. But, as
$\Delta v$ grows, and especially when $\Delta v > U$, the single excitation energy curve approaches its KS alterego, but 
the double excitation does not.  This is because, in the charge transfer regime when $\Delta v > U>> 2t$, both electrons are on the same site for the ground state (e.g. site 1), on opposite sites for the first excitation (and also not interacting),  whereas for the double, both electrons are again one site (site 2).  The reverse is true for the Mott-Hubbard regime, defined for the region of $U>> \Delta v>>2t$ 

Next we consider TDDFT within the adiabatically exact approximation (AEA).  
The extremely small Hilbert space means the response function is not a matrix but a
single function \cite{CFMB18} that vanishes at each excitation:
\ben
\chi^{-1}(\omega)=\chi\s^{-1}(\omega)-f\Hxc(\omega).
\een
If $f\Hxc$ is ignored, transitions occur at $\omega=\omega_{s}$.
If $f\Hxc(\omega=0)$ is used (AEA), it shifts the positions of the single excitations, but still misses all higher excitations. This is the best possible
performance of the adiabatic approximation, because we used the exact ground state functional to determine $f\Hxc (\omega=0)$. This produces
the green curve for the single excitation in Fig. (\ref{boat:2vAE_short}).  We see that AEA is extremely accurate (because correlation
is weak), and becomes even more so as the asymmetry is increased.  But there is no analogous
curve for the double excitation, as there is no way to access the double within linear-response TDDFT
without a frequency dependent kernel.  (Even higher-order perturbation theory can at most yield 
doubles that are twice the singles, which would not be accurate \cite{CM00}).

\begin{figure}[!htp]
  \includegraphics[width=\linewidth]{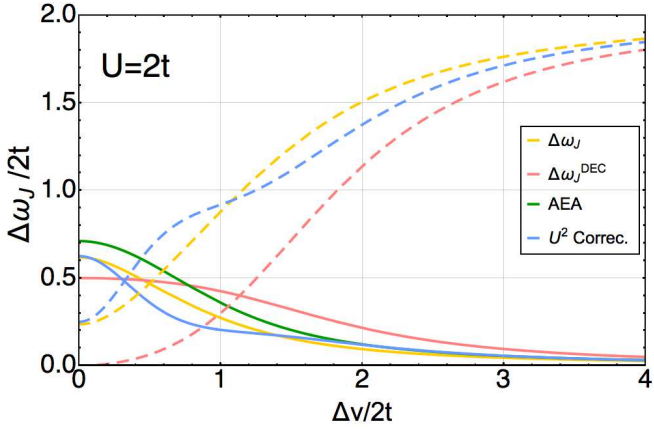}
  \caption{Correction to the KS transitions of Fig. \ref{boat:2vAE_short}, both 
exact and various approximations, where solid lines are single excitations, dashed are double.
The correction to the single turns off with increasing asymmetry, but not so the double excitation.
The DEC/EEXX approximation correctly captures both effects.  Also included is the leading
correlation contribution, which further improves the results, when the system is weakly correlated.}
  \label{dw1}
\end{figure}
Next we apply EDFT to the dimer.  The results for the single are well-known \cite{DMF17}, because
they can be extracted from a bi-ensemble of the ground and first excited states, and there is no difference between the GOK and GOK II ensembles. But to extract the double excitation we use a three state GOK II ensemble. Applying Eq.(3) to the KS eigenstates of the Hubbard dimer, one finds:
\ben 
\Delta \omega^{DEC}_{1} = \frac{U}{2}\bigg(1- \frac{\Delta n^{2}}{4}\bigg),~~~
\Delta \omega^{DEC}_{2}= \frac{U}{2}\Delta n^{2},
\een
which agree perfectly with Eq.(\ref{eq.1}), applied
to the dimer and expressed in terms of the ground-state density \cite{CFMB18}.  These yield the red lines in Fig.\ref{boat:2vAE_short}. 

To analyze and expand on these results, in Fig. \ref{dw1} we directly plot 
$\Delta \omega_{J}$ for each transition.  This is the true measure of the quality
of an approximate treatment of excitations, as the KS transitions are 
determined entirely by ground-state DFT.   We use the single as a test
case, as the analytic results are already known.  
The DEC/EEXX curves are comparable to those of the AEA 
TDDFT, doing better for $\Delta v < 2t$, but worse as the asymmetry increases, similar 
to its performance for both atoms and the Hooke's atom \cite{YPBU17}.
As $\Delta v\to \infty$, $\Delta n \to 2$, turning off the corrections to the single.

Now we focus on the main interest, the double excitation.
Here DEC/EEXX yields no
correction at $\Delta v=0$, but everywhere else reduces the error of the KS transition, but
with substantially greater error than for the single.  This is consistent with the earlier
results, but can we discern here if this is accidental or not?
To do this, we take advantage of the model's simplicity, and the many results that are already known.
One peculiarity is that, performing a many-body expansion for fixed
$\Delta v$ as a function of $U$, one finds that the double excitation has no first-order correction, i.e.,
correction to the tight-binding result is of order $U^2$ \cite{CFMB18}.  This would appear to make it useless for
our purposes.  However, $\omega_{s}$, by virtue of its dependence on the ground-state
density, {\em does} have a first-order correction in $\lambda$, which means that $\Delta\omega_2$
is also first-order, and is
correctly captured by the DEC/EEXX approximation, as shown.   This correction happens to vanish
at $\Delta v=0$.  (This means that studying only the symmetric dimer would produce qualitatively
incorrect conclusions on this point.)

Because of the simplicity of the model, we can use the results of Ref.\cite{CFMB18} to derive the next
correction in powers of $U$ (or $\lambda$), by converting $\Delta v$-dependence to $\Delta n$-dependence, yielding
\bea
\Delta \omega^{(2)}_{1}&= &\frac{\sqrt{4-\Delta n^2}(4-13\Delta n^{2} +3\Delta n^{4})U^{2}}{64(2t)},\nonumber\\
\Delta \omega^{(2)}_{2}&=& \frac{\sqrt{4-\Delta n^2}(4+11\Delta n^{2} -3\Delta n^{4})U^{2}}{32(2t)}.
\eea
Note that these corrections {\em cannot} be deduced from the DEC/EEXX of Ref. \cite{YPBU17}.  These are
shown in Fig. \ref{dw1} and (almost) everywhere reduce the error of DEC, as expected in the
weakly correlated regime.  Moreover, they do produce great improvement in the double at $\Delta v=0$, and
so provide a benchmark for correlation corrections to DEC/EEXX.

\begin{figure}[!htp]
  \includegraphics[width=\linewidth]{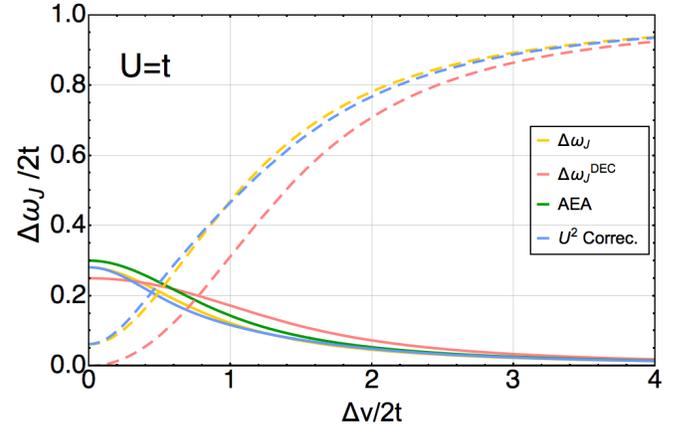}
  \caption{Same as Fig. \ref{dw1} but at a weaker correlation of $U= t$, 
showing that the DEC improves as correlation weakens, and the
second order correction agrees even better. }
  \label{dw05}
\end{figure}
To make sure our understanding is correct, in Fig. \ref{dw05} we show the results when $U=t$,
i.e. the same system but with weaker correlation.  Now the second-order correction is almost
perfect everywhere, showing perturbation theory is converging.  Moreover, the absolute errors in 
DEC have halved, but remain large out to about $\Delta v=2U$.
\begin{figure}[!htp]
  \includegraphics[width=\linewidth]{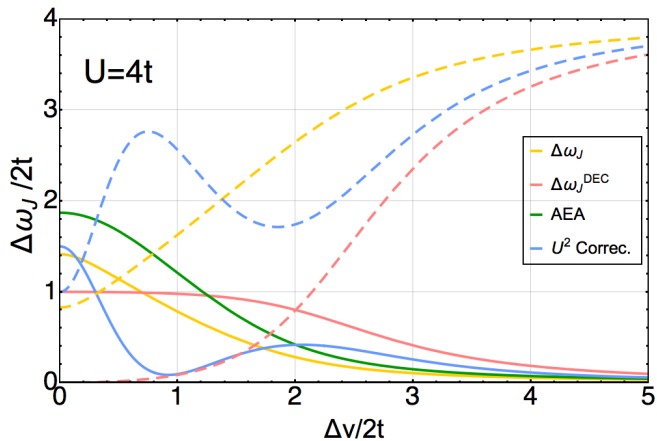}
  \caption{Same as Fig. \ref{dw1}, but at a stronger correlation of $U= 4t$, showing the
failure of DEC when correlation is strong. 
Here DEC fails for small $\Delta v$, but nonetheless agrees
for both DEC and its second-order correction 
for large values of $\Delta v$, showing that the system becomes weakly correlated when
asymmetry dominates over correlation strength.}
  \label{dw2}
\end{figure}
In our last figure, Fig. \ref{dw2}, we show what happens as GL perturbation theory begins to fail.
Near $\Delta v=0$, DEC fails completely, with equal corrections to the first and second excitation, making
the gap precisely zero.  This is where convergence of perturbation theory breaks down, and the
KS transitions are not a good starting point.  However,
even here, for $\Delta v >> U$, the gap is much larger, and both single and double corrections become
accurate.  This is consistent with the claims of Ref.\cite{CFMB18}, that for $\Delta v > U$, a system is always
weakly correlated, no matter how large $U$ is, as far as DFT is concerned.

{\bf Discussion:}  So what can we conclude from this very simple model?  The most important thing is
that, generically, the DEC/EEXX approximation yields a meaningful and non-zero correction to
a double excitation, producing the exact linear term in the GL perturbation for fixed density.
In special cases where this term vanishes identically, it is of course useless.  In some ways, our
case is more typical than either of those studied in Ref.\cite{YPBU17}, as all cases studied there involved
double excitations in regions of the energy spectrum with single excitations nearby (where dressed
TDDFT could be applied), but here we have a double excitation without a single nearby (and hence
dressed TDDFT would not work).  

How would TDDFT capture these effects if correlation is included?  Ref.\cite{CFMB18} gives the answer
for this model.  There is a pole in the kernel that generates the double excitation.  It has the
form:
\ben
f{\Hxc} (\omega) \approx \frac{a}{\omega-\omega_p}.
\een
Now, both the numerator and $\omega_p$ have expansions in powers of $\lambda$.  While $\omega_p$
contains all orders, $a$ starts at second-order.  It is meaningless to speak of an expansion of
the kernel in powers of $\lambda$, as this expansion always fails as $\omega$ approaches the pole.
Both the numerator and the transition frequencies have well-behaved expansions in powers of $\lambda$, and can be usefully approximated in a power series when the system is weakly correlated, but the kernel in TDDFT never does.

Although the Hubbard dimer is not a quantitative model of anything, it roughly approximates a minimal 
basis model for a diatomic with two valence electrons.  In the symmetric case, this would correspond to 
H$_2$.  As the bond is stretched, $t\to 0$, but $U$ and $\Delta v$ remain finite, so $U/(2t)$ and $\Delta v/(2t)\to \infty$.
For H$_2$, by symmetry, $\Delta v=0$, and this may present special difficulties for DEC/EEXX, as the linear
contribution might be unusually small.  On the other hand, for LiH, it should work well.

Finally, while this model may appear overly simple, its great power lies in the ability to show
transparently what is going on.  It clearly demonstrates that EDFT \textit{can} accurately capture double excitations, even when using an EEXX approximation, with no correlation. It would be highly non-trivial (and time consuming) to perform all these TDDFT and 
EDFT calculations on more realistic systems, and impossible to write down and examine the behavior
of analytic expressions.  A longer paper on the subject is in preparation.

{\bf Acknowledgements:} F.S. and K.B kindly acknowledge support from NSF grant CHE- 1464795.  

\bibliography{Master_DEC}

\label{page:end}
\end{document}